\documentclass[aps,superscriptaddress,showpacs]{revtex4}
\usepackage{graphicx}
\usepackage{subfigure}
\usepackage{bm}
\begin{document}
\title{Stretching of polymers around the Kolmogorov scale in 
       a turbulent shear flow}
\author{Jahanshah Davoudi\footnote{Present address: International Centre for Theoretical Physics,
34014 Trieste, Italy}}
\affiliation{Department of Physics, Philipps University Marburg, 
           D-35032 Marburg, Germany}
\author{J\"org Schumacher\footnote{Present address: Department of Mechanical Engineering, Ilmenau
University of Technology, D-98684 Ilmenau, Germany}}
\affiliation{Department of Physics, Philipps University Marburg, 
           D-35032 Marburg, Germany}
\date{\today}
\begin{abstract}
We present numerical studies of stretching of Hookean dumbbells in a
turbulent Navier-Stokes flow with a linear mean profile, $\langle u_x\rangle=Sy$. 
In addition to the turbulence features beyond the viscous Kolmogorov scale $\eta$, 
the dynamics at the equilibrium extension of the dumbbells significantly below
$\eta$ is well resolved. The variation of the constant shear rate $S$ 
causes a change of the turbulent velocity fluctuations on all scales and 
thus of the intensity of local stretching rate of the advecting flow. 
The latter is measured by the maximum Lyapunov exponent $\lambda_1$ which 
is found to increase as $\lambda_1\sim S^{3/2}$, in agreement with a dimensional 
argument.
The ensemble of up to $2\times 10^6$ passively advected dumbbells is advanced by 
Brownian dynamics simulations in combination with a pseudospectral integration
for the turbulent shear flow. Anisotropy of stretching is quantified by the 
statistics of the azimuthal angle $\phi$ which measures the alignment with the 
mean flow axis in the $x$-$y$ shear plane, and the polar angle $\theta$ which 
determines the orientation with respect to the shear plane. The asymmetry of the
probability density function (PDF) of $\phi$ increases with growing shear rate $S$. 
Furthermore, the PDF becomes increasingly peaked around mean flow direction 
($\phi= 0$). In contrast, the PDF of the polar angle $\theta$ is symmetric and less 
sensitive to changes of $S$.
\end{abstract}
\maketitle

\section{Introduction}
When a few parts per million in weight of long-chained polymers are added to a 
turbulent fluid its properties change drastically 
and a significant reduction of turbulent drag 
is observed (see e.g. Refs.~1 and 2 for 
reviews). Although the phenomenon is known from pipe flow experiments
for almost 60 years \cite{Toms1949} a complete 
understanding is still lacking. 
It is observed in wall-bounded flows \cite{Warholic1999,White2000,White2003} as well as in bulk 
turbulence.\cite{Wagner2003,Liberzon2005}
The macroscopic drag reduction seems to go in line 
with a microscopic transition of the macromolecules from a preferentially 
coiled near-the-equilibrium to a stretched non-equilibrium state.
The relation of the polymer dynamics at small scales to the flow
statistics at larger scales is therefore essential for solving the 
drag reduction problem.\cite{Liberzon2005,Terrapon2004}

The dynamics of single long polymer chains in simple and partly steady flows
is already quite complicated and rich with respect to conformational variety. This was 
demonstrated
in experiments and Brownian dynamics simulations. 
\cite{Perkins1997,Smith1999,Hur2001,Graham2003,Puliafito2005} 
Only recently, a more complex  situation was investigated --  
the stretching of polymer chains in random isotropic and random linear shear flows --
by means of experiments in the so-called elastic turbulence limit.\cite{Steinberg2001,Steinberg2005}
Analytic solutions are possible for the kinematic stretching of polymers in isotropic and 
white-in-time Kraichnan flow.\cite{Balkovsky2000,Thiffeault2003,Celani2005,Vincenzi2005} 
The polymer chain is 
there described as two beads that are connected by a spring. The model is known as the
mesoscopic dumbbell model and the entropic spring force follows either
a linear Hookean or a nonlinear law. The latter captures the finite 
extensibility of the polymers.\cite{Bird1987} 
Qualitatively new effects can be observed in a shear flow such as
the tumbling, a flipping of the chain in the plane that is spanned by the 
shear flow.\cite{Smith1999,Schroeder2005}
Analytical predictions for the tumbling statistics of dumbbells are possible when the  
isotropic Kraichnan flow is superposed with a linear shear flow at a very large shear 
rate.\cite{Chertkov2005} The dynamics of the angular
degrees of freedom can then be simplified significantly because the large shear aligns
the dumbbells preferentially with the mean flow direction. 

Although these studies give us a lot of helpful insights about the statistics of the extension and
orientation of the polymer chains, the situation in a turbulent Navier-Stokes fluid
is more complex. The flow structures are correlated in space {\it and} time and it
can be expected that they cause significant differences in the stretching history 
of a chain while moving through the flow. Furthermore, turbulent fluctuations and the
large scale shear are intimately coupled to each other. In other words, one cannot 
consider them as independently adjustable parameters. 
An increase of the shear rate $S$ will change the Reynolds number of the flow and more 
importantly the ratio and
magnitude of the turbulent velocity fluctuations $\langle u_i^2\rangle$ with $i=x,y,z$
(see Refs.~24 and 25).
For example, the presence of shear causes the generation of
streamwise streaks, that enlarge the fluctuations in the streamwise direction.\cite{Robinson1991} 
As a consequence, we find that the maximum stretching rate grows
with respect to $S$ faster as recently predicted.\cite{Chertkov2005}

The present work consists of two parts. 
The first part will focus on the statistics of the flow itself, at scales below and 
above the viscous Kolmogorov scale, $ \eta=\frac{\nu^{3/4}}{\langle\epsilon\rangle^{1/4}}$,
with $\nu$ being the kinematic viscosity and $\langle\epsilon\rangle$ 
the mean energy dissipation
rate of the flow. We consider a simple turbulent shear flow with a linear mean profile
which allows for studying the effects of shear on Lagrangian stretching rates. This in turn
will reveal a different regime of polymer stretching in comparison 
with an analytic model of Chertkov {\it et al.}\cite{Chertkov2005}
The second part presents studies of the kinematics of polymer stretching in such
flows. We will demonstrate that the interplay between the shear rate $S$ and the Weissenberg number
$Wi$ is important for the stretching statistics. The dimensionless parameter $Wi$ relates the 
local stretching rate of the flow to the relaxation rate of the polymers. Only the simplest 
mesoscopic polymer model,  
the Hookean dumbbell model \cite{Bird1987}, is discussed here. In the present study, the
dumbbells do not react back on the shear flow. As it will turn out, turbulence 
is able to stop the stretching of the linear springs, but at a scale 
that is larger than the Kolmogorov scale of the flow.\cite{Tabor1986} Our approach
intends  
to keep the polymer dynamics simple, but to take an advecting flow in its full
turbulent complexity stemming from the Navier-Stokes equations. 

In chapter II we present the dynamical equations for the flow and the dumbbells and 
discuss characteristic parameters. The dependence of local stretching rates 
on the shear rate is quantified in chapter III. The subsequent chapter summarizes 
our findings on the dumbbell dynamics where extension and angular statistics are discussed.
Eventually, a summary and an outlook are given.

\section{Model and equations}

\subsection{Advecting fluid} 
The Navier-Stokes equations for a three-dimensional incompressible fluid
are solved by a
pseudo-spectral method using a second-order predictor-corrector scheme for advancement
in time.\cite{Yeung1988} The equations of motion are 
%-------------------------------------------------------------------------------
\begin{eqnarray}
\label{nseq}
\frac{\partial{\bf u}}{\partial t}+({\bf u}\cdot{\bf \nabla}){\bf u}
&=&-{\bf \nabla} p+\nu {\bf \nabla}^2{\bf u}+{\bf f}\,,\\
\label{ceq}
{\bf \nabla}\cdot{\bf u}&=&0
\end{eqnarray}
%-------------------------------------------------------------------------------
where ${\bf u}$ is the (total) velocity field, $p$ the kinematic pressure field, and ${\bf f}$ 
the volume force density.
The case without shear is solved in a three-dimensional 
box of side-length $L=2\pi$ with periodic boundary conditions.
The nearly homogeneous shear flow is modeled in a volume with free-slip boundary 
conditions in the shear direction $y$ and periodic boundaries otherwise. 
Here the total velocity field follows by a Reynolds (de)composition as a linear mean part
with the constant shear rate $S$ and a turbulent fluctuating part
%--------------------------------------------------------------------------------
\begin{equation}
{\bf u}=\langle{\bf u}\rangle + {\bf v}=Sy{\bf e}_x+{\bf v}\,.
\label{Reynolds}
\end{equation}
%--------------------------------------------------------------------------------
The aspect 
ratio is $L_x :L_y :L_z=2\pi:\pi:2\pi$. The applied volume forcing is a combination
of an isotropic forcing that injects energy at a fixed rate of $\epsilon_{in}=0.1$
and a shear forcing. Consequently, $\langle\epsilon\rangle\equiv\epsilon_{in}$ for the
statistically stationary case at $S=0$.
More details on the volume forcing and the numerical scheme can be found in Ref.~29.
In Tab.~1, we summarize some statistical properties of the flows that were studied. 
Since we want to resolve eventually a range of scales below $\eta$, we are limited 
in the range of accessible Reynolds numbers although resolutions of up to 
$512\times 257\times 512$ grid
points are used for the simulations.
%-----------------------------------------------------------------------
\renewcommand{\arraystretch}{1.3}
\begin{table}
\begin{center}
\begin{tabular}{lcccccc}
\hline\hline
run          & 1    & 2       & 3       & 4       & 5       & 6  \\ \hline
$S$          & 0    & $1/\pi$ & $3/\pi$ & $5/\pi$ & $7/\pi$ & $9/\pi$ \\
$\nu$        & 1/30 & 1/30    & 1/30    &  1/30   & 1/30    & 1/30 \\ 
$\langle\epsilon\rangle$ & 0.100 & 0.113 & 0.285  & 0.845 & 2.268 & 5.179\\ 
$R_{\lambda}$& 10.5 & 9.1     & 19.2    &  29.8   & 36.2  & 43.7   \\ 
$S^*$        & 0    & 1.21    & 4.85    &  7.31   & 7.58  & 7.78   \\ 
$S\tau_{\eta}$ & 0  & 0.17    & 0.33    &  0.32   & 0.27  & 0.23   \\
$T_{av}/T$ & 20.5 & 25.1    & 18.9    &  21.0   & 112.8 & 70.7   \\ 
$k_{max}\eta$& 8.39 & 8.12    & 6.44    &  4.91   & 3.84  & 3.12   \\ \hline\hline
\end{tabular}
\caption{Parameters of the direct numerical simulations. $S$ is the 
constant mean shear rate, $\nu$ the kinematic viscosity, $\langle\epsilon\rangle$ 
is the mean energy dissipation rate, 
$R_{\lambda}=\sqrt{5/(3\langle\epsilon\rangle\nu)} \langle v^2\rangle$ is
the Taylor microscale Reynolds number with the mean square of the turbulent 
velocity fluctuations $\langle v^2\rangle$, and 
$S^*=S\langle v^2\rangle/\langle\epsilon\rangle$ is the dimensionless shear parameter.
The averaging time $T_{av}$ is given in units of the large scale eddy turnover time
$T=\langle v^2\rangle/(2\langle\epsilon\rangle)$.
The spectral resolution is indicated by $k_{max}\eta$ where $k_{max}=\sqrt{2}N/3$. 
All studies of the flow properties were done with $N=128$.}
\end{center}
\label{tab1}
\end{table}

%-----------------------------------------------------------------------

\subsection{Hookean dumbbells}
The simplest description of polymer stretching can be accomplished by considering
Hookean dumbbells. Their entropic elastic force is linearly 
dependent on the separation vector,
${\bf R}={\bf x}_2-{\bf x}_1$, that is spanned between both beads at positions 
${\bf x}_2(t)$ and ${\bf x}_1(t)$, respectively. When taking into account the elastic 
entropic force, 
hydrodynamic Stokes drag, and thermal noise the evolution equation for ${\bf R}$ reads 
\cite{Bird1987,Oettinger}
%----------------------------------------------------------------------------------
\begin{equation}
\dot{\bf R}=\Delta{\bf u}-\frac{{\bf R}}{2\tau}+\sqrt{\frac{R_0^2}{\tau}}{\bf \xi},
\label{deq}
\end{equation}
%----------------------------------------------------------------------------------
where $\Delta {\bf u}= {\bf u}({\bf x_2},t)-{\bf u}({\bf x_1},t)$ is the relative fluid velocity  
at the bead centers. The last term is the thermal Gaussian noise with the following properties
%----------------------------------------------------------------------------------
\begin{eqnarray}
\langle\xi_i(t)\rangle&=&0\,\\
\langle\xi_i(t)\xi_j(t^{\prime}\rangle&=&\delta_{ij}\delta(t-t^{\prime})\,
\label{white}
\end{eqnarray}
%----------------------------------------------------------------------------------
with $i,j=x,y,z$. It prevents 
the extension of a dumbbell to shrink below its equilibrium length 
%----------------------------------------------------------------------------------
\begin{eqnarray}
R_0=\sqrt{\frac{k_B T}{H}}\,,
\label{equilibrium}
\end{eqnarray}
%----------------------------------------------------------------------------------
with $k_B$ being the Boltzmann constant, $T$ the temperature, and
$H$ the spring constant of the Hookean spring. Equation (\ref{equilibrium})
follows from the equipartition
theorem.  Time constant $\tau$ is the relaxation time of the polymers and is given for 
dumbbells as \cite{Bird1987}
\begin{equation}
\tau=\frac{\zeta}{4H},
\end{equation}
where $\zeta=6\pi\rho\nu a$ is the Stokes drag coefficient with fluid mass density $\rho$ and
bead radius $a$. We sketch in Fig.~1 the coordinate system that is used.
Since we restrict our study to Hookean dumbbells, the effects of finite 
extension are not imposed by the potential. Therefore the extended dumbbells experience 
both the smooth and rough scales of the advecting flow during their time evolution. 
Consequently, the velocity difference 
$\Delta {\bf u}$
is {\it not} approximated by the linearization $({\bf R\cdot \nabla}){\bf u}$ 
in Eq.~(\ref{deq}) as usually done. We solve the 
equations for the beads separately, but can reconstruct the full dynamics of ${\bf R}$ 
afterwards. 
An ensemble of up to $2\times 10^6$ dumbbells, i.e., $4\times 10^6$ beads, is advanced by a 
weak second-order predictor-corrector scheme simultaneously with the flow 
equations.\cite{Oettinger}
Initially, their center of mass is seeded randomly in space with a uniform distribution and an 
initial extension of $R_0$.
All Lagrangian interpolations were done with a trilinear scheme since the velocity field is very 
well resolved as given by the spectral resolution factor $k_{max}\eta$ in Tab.~1.
%-------------------------------------------------------------------------------
\begin{figure}[htb]
\centerline{\includegraphics[angle=0,scale=0.5,draft=false]{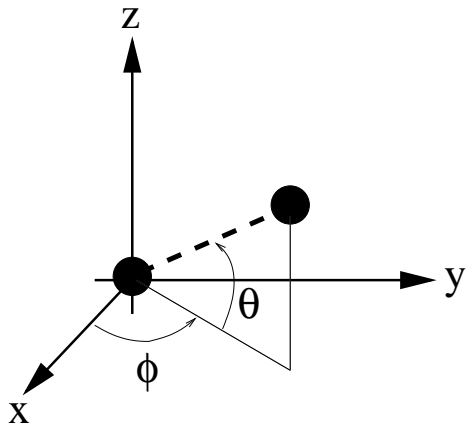}}
\caption{Dumbbell coordinate system that is used throughout this text.
It follows from the sketch that $R_x=R\cos\phi\cos\theta$, $R_y=R\sin\phi\cos\theta$, 
and $R_z=\sin\theta$
where $R$ is the distance between both beads.
The notation differs from conventional spherical coordinates, but has the advantage of giving
alignment with the outer mean flow for $\phi=\theta=0$. $\phi$ is azimuthal angle and $\theta$
the polar angle.}
\label{geometry}
\end{figure}
%---------------------------------------------------------------------------------------

\subsection{Stretching rates}
According to Eq.~(\ref{deq}) the dynamics of dumbbells is subject to both, local
stretching rate due to varying strain at small scales and restoring linear spring force.
The separation vector between two fluid elements, $\left |\delta {\bf r} \right|$, evolves as
%-------------------------------------------------------------------------------
\begin{equation}
\frac{\mbox{d}}{\mbox{d}t}\delta r_j(t)=\sigma_{jk}(t)\,\delta r_k(t)\;\;\;\mbox{for}\;\;\;j,k=x,y,z\,,  
\end{equation}
%-------------------------------------------------------------------------------
where $\sigma_{jk}(t)$ is the local stretching tensor along the Lagrangian trajectories. 
Consequently,
%--------------------------------------------------------------------------------
\begin{equation}
\delta r_j(t)=W_{jk}(t,0)\delta r_k(0)\,,
\end{equation}
%--------------------------------------------------------------------------------
with the time-ordered exponential
%-------------------------------------------------------------------------------
\begin{equation}
W_{jk}(t,0)={\cal T}_{+}\exp\left(\int_0^t\,\sigma_{jk}(\tau)\,\mbox{d}\tau\right)\,,  
\end{equation}
%-------------------------------------------------------------------------------
where $\delta{\bf r}(0)$ is the initial separation vector. 
The local stretching and contraction rates are quantified by the Lyapunov
exponents 
%-------------------------------------------------------------------------------
\begin{equation}
\lambda_i=\lim_{t\to\infty}\frac{1}{t}\langle\log(\frac{|\delta {\bf r}^{(i)}(t)|}
{|\delta {\bf r}^{(i)}(0)|})\rangle_L\,,
\end{equation}
%-------------------------------------------------------------------------------
with $i=1,2,3$ for three dimensions and 
$\lambda_1\ge\lambda_2\ge\lambda_3$.\cite{Benettin1980,Girimaji1988} 
The subscript $L$ is an average over different realizations of the Lagrangian tracer 
tracks. The largest Lyapunov exponent, $\lambda_1$, quantifies
the growth of the norm of the separation vector $\delta{\bf r}$. 
The exponents are long-time averages
of Lagrangian simulations following the algorithm 
by Benettin {\it et al.} \cite{Benettin1980}. Figure~2 illustrates that the convergence of 
the Lyapunov spectrum for a larger shear 
rate requires a long time series. A sufficient convergence is reached there after about
70 large eddy turnover times the latter of which is defined as 
$T=\langle v^2\rangle/(2\langle\epsilon\rangle)$. 
Clearly this procedure has to be done for each $S$ value anew.  

While the mean stretching time scale in a turbulent flow is given by the $\lambda_1^{-1}$, the beads 
relax back to their coiled equilibrium distance within times
of the order of $\tau$. The Weissenberg number which is defined as
%----------------------------------------------------------------------------
\begin{equation}
Wi=\lambda_1 \tau\,,
\end{equation}
%--------------------------------------------------------------------------
is an appropriate measure for the competition of those mechanisms.
For $Wi<1/2$ the relaxation to equilibrium size is on average faster than 
stretching by the flow. Polymers are preferentially in a coiled state and their size
distribution is stationary. In contrast, for $Wi>1/2$ stretching proceeds in time until the
finite extensibility limit is reached or turbulence stops the growth of dumbbells. This
transition at $Wi\simeq 1/2$ is known as the coil-stretch transition.\cite{deGennes1974}
The properties of such a transition in random and short-in-time correlated flows have 
been studied analytically.\cite{Balkovsky2000,Thiffeault2003,Celani2005,Vincenzi2005}
The probability density function (PDF) of end-to-end distance $R=|{\bf R}|$ exhibits then
algebraic tails for $Wi<1/2$, i.e. when $R \to \infty$ one gets a scaling law
$P(R) \sim R^{-1-q}$ where $q$ depends linearly on $Wi^{-1}$. In case of Hookean dumbbells, 
the PDF is not any longer normalizable at values $Wi>1/2$. 

Although the analytical treatment is possible only for this limited
class of random short-correlated flows, the coil-stretch transition appears in other turbulent
flows. Experimental \cite{Wagner2003,Steinberg2001} and several numerical studies
\cite{Terrapon2004,Graham2003,Celani2005,Sureshkumar1997,Eckhardt02,Ilg2002,Vaithianathan2003} 
all provide direct or indirect verifications of such a transition within different models 
for polymers or in distinct flow geometries. A sharp transition from a preferentially
coiled to a preferentially stretched state at $Wi=Wi_c$ is however often not observable since different 
local flow topologies cause differently large mean extensions, as known from single molecule
experiments.\cite{Perkins1997,Smith1999}     
%---------------------------------------------------------------------------------------
\begin{figure}[htb]
\centerline{\includegraphics[angle=0,scale=0.6,draft=false]{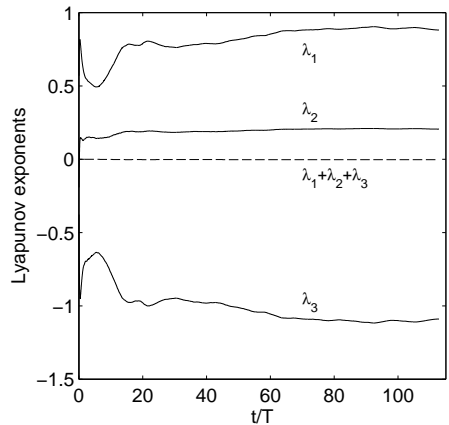}}
\caption{Time evolution of the three Lyapunov exponents ${\lambda}_i$ 
for the simulation with a shear rate $S=7/\pi$. The dashed line is the sum of the
exponents which has to be zero due to incompressibility of the flow. Time is in units
of the large scale eddy turnover time 
$T=\langle v^2\rangle/(2\langle\epsilon\rangle)$. 1000 Lagrangian tracer particles were
seeded homogeneously at the beginning of the simulation.}
\label{lyap}
\end{figure}
%---------------------------------------------------------------------------------------
\begin{figure}[htb]
\centerline{\includegraphics[angle=0,scale=0.6,draft=false]{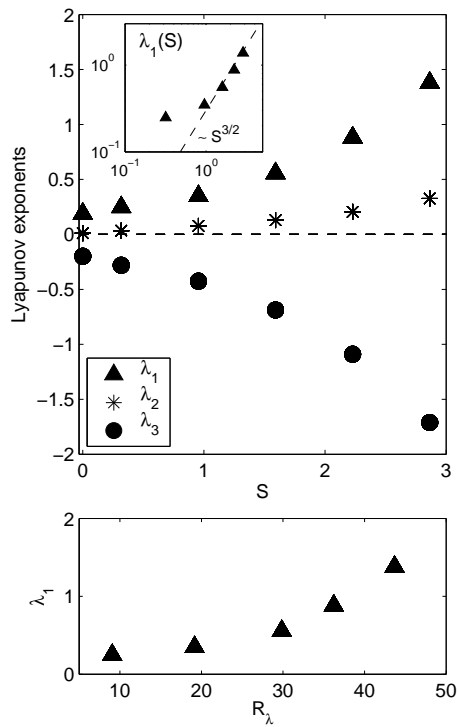}}
\caption{Upper panel: Lyapunov spectrum $\lambda_i$ for $i=1,2,3$
as a function of
shear rate $S$. The inset replots $\lambda_1(S)$ in logarithmic axes and fits a power law
$\lambda_1\sim S^{3/2}$ to the larger shear rate values. 
Lower panel: Largest Lyapunov exponent $\lambda_1$ as a 
function of the Taylor microscale Reynolds number, 
$R_{\lambda}=\sqrt{5/(3\langle\epsilon\rangle\nu)} \langle v^2\rangle$.}
\label{lyapshear}
\end{figure}
%---------------------------------------------------------------------------------------

\section{Stretching properties of the shear flow}
Let us discuss now the stretching properties of the shear flow.
The dependence of the Lyapunov spectrum on the shear rate $S$ is shown in 
upper panel of Fig.~3. 
One can see that all three exponents grow monotonically in magnitude
with increasing $S$.  The lower panel of the same
figure shows the largest Lyapunov exponent as a function of the Taylor microscale 
Reynolds number $R_{\lambda}$. Furthermore, the strength of the shear can be quantified by the 
dimensionless shear parameter $S^*=S\langle v^2\rangle/\langle\epsilon\rangle$. 
Table~1 indicates that this parameter saturates with growing shear rate to a value of about 8 while 
the Reynolds number is still
increasing with growing shear rate. Such a saturation of $S^*$ is known from studies in homogenous
shear flows in the stationary regime. \cite{Schumacher2004,Gualtieri2002,Yakhot2003} 
Our findings as given by Fig.~3 and Tab.~1 underline the dependence of 
all turbulence parameters on the shear rate $S$.  

An analytic treatment has to rely on a simpler 
case. In Ref.~23, the fluctuating Kraichnan flow is added with a linear
shear flow which gives a decomposition as in (\ref{Reynolds}). The correlation matrix
of the Gaussian fluctuating strain is defined as 
$\langle \nabla_i v_j(t)\nabla_k v_l(t^{\prime})\rangle=D C_{ijkl} \delta(t-t^{\prime})$.
It contains an amplitude parameter $D$ that models the strength of the small-scale 
fluctuations and 
can be varied {\it independently} beside the shear rate $S$ in the linear part. 
The amplitude $D$ introduces a fluctuation time scale $T_f=D^{-1}$ beside the shear time scale 
$T_s=S^{-1}$. 

In the large shear limit, which is defined as
%-----------------------------------------------------------------------------------------
\begin{equation}
T_s=\frac{1}{S}\ll T_f=\frac{1}{D}\,,
\label{chertkov1}
\end{equation}
%-----------------------------------------------------------------------------------------
the molecules are most of the time aligned with the 
shear. The azimuthal angle $\phi$ (see Fig.~1) has a nonzero average 
$\langle\phi\rangle\simeq (D/S)^{1/3}$
and satisfies the asymptotic power law distribution $P(\phi) \sim 1/\phi^{2}$  for an intermediate
range of $\langle\phi\rangle\ll\phi\ll 1$.  
The positive Lyapunov exponent follows to $\lambda_1 \sim S \langle \phi \rangle$ and thus 
%-----------------------------------------------------------------------------------------
\begin{equation}
\lambda_1 \sim S^{2/3}\,.
\label{chertkov2}
\end{equation}
%-----------------------------------------------------------------------------------------
Relation (\ref{chertkov1}) is known in Navier-Stokes turbulence as being close to
the rapid distortion limit of a shear flow \cite{Townsend1956}, when the large shear rate 
determines the turbulent dynamics. 

There is a slight increase of $\lambda_1$ in comparison to the isotropic case for
$S\leq 1$ as the inset of the Fig.~3
indicates. However, as the shear rate increases further we observe a crossover to a faster
nonlinear increase with $S$. 
In fact $\lambda_1 \sim 
\sqrt{\langle \epsilon \rangle /\nu}$ and $\langle \epsilon \rangle \sim S^3 L^2$ leads to 
the $3/2$ scaling,
%-----------------------------------------------------------------------------------------
\begin{equation}
\lambda_1\sim S^{3/2}\,,
\label{dav1}
\end{equation}
%-----------------------------------------------------------------------------------------
which differs from (\ref{chertkov2}). Although the range of our shear rates is limited,
the observed scaling is consistent with this dimensional estimate.
This argument and the saturation of the dimensionless shear parameter suggest
that our shear flow operates in a different regime. Let us therefore compare shear and fluctuation
time scale in the present case. The small-scale gradient fluctuations are captured by the
Kolmogorov time $\tau_{\eta}=\sqrt{\nu/\langle\epsilon\rangle}$. $T_s$ has to be related 
to $\tau_{\eta}$ which is known as the Corrsin parameter.\cite{Corrsin1958}
Figure~4 shows that 
%-----------------------------------------------------------------------------------------
\begin{equation}
T_s\gtrsim \tau_{\eta}=\sqrt{\frac{\nu}{\langle\epsilon\rangle}}\;\;\mbox{or}\;\;S\tau_{\eta}\lesssim 1\,.
\label{dav2}
\end{equation}
%-----------------------------------------------------------------------------------------
In fact (\ref{dav1}) dictates $S \tau_{\eta} \sim S^{-1/2}$.
In order to support this behavior we add  
Tab.~2 where data from DNS at higher Reynolds number \cite{Schumacher2004} and from
turbulent boundary layer measurements \cite{Knobloch2004,Jachens2004} are shown. The 
experimental data demonstrate that even close to the wall the
Corrsin parameter does not exceed unity as can be seen for the data at $y^+=34$.
We conclude that our studies of the polymer dynamics in the turbulent
shear flow cannot be compared with the analytic results of Chertkov {\it et al.} \cite{Chertkov2005}
%---------------------------------------------------------------------------------------
\begin{figure}[htb]
\centerline{\includegraphics[angle=0,scale=0.6,draft=false]{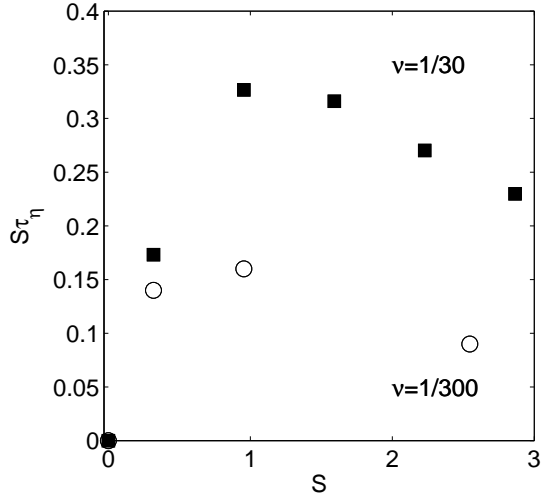}}
\caption{Ratio of the fluctuation time scale to the shear time scale,
$S\tau_{\eta}$. Present data are filled symbols and the open symbols are taken from
Ref.~29 (see also Tab.~2).}
\label{shear}
\end{figure}

%-----------------------------------------------------------------------
\renewcommand{\arraystretch}{1.3}
\begin{table}
\begin{center}
\begin{tabular}{lcccccc}
\hline\hline
       & $\nu$ & $S$     & $y^+$ & $R_{\lambda}$ & $S^*$ & $S\tau_{\eta}$ \\ \hline
DNS1   & 1/300 & $1/\pi$ & - & 55 & 4.2 &  0.14\\
DNS2   & 1/300 & $3/\pi$ & - & 175& 7.7 &  0.16\\ 
DNS3   & 1/300 & $8/\pi$ & - & 309& 9.8 &  0.09\\ \hline 
HFI1   & $\nu_{air}$ & 1576 & 34   &  151 & 37.8 & 0.97   \\ 
HFI2   & $\nu_{air}$ & 20   & 1549 &   75 &  2.8 & 0.14   \\ \hline
DNW1   & $\nu_{air}$ & 1447 & 709  &  1156& 50.3 & 0.17   \\ 
DNW2   & $\nu_{air}$ & 151  & 7826 &  1739& 20.7 & 0.05   \\ 
DNW3   & $\nu_{air}$ & 78   & 22753&  1299& 14.3 & 0.04   \\ \hline\hline
\end{tabular}
\caption{Further data on the ratio $S\tau_{\eta}$ taken from Ref.~29
(see series II in Tab.~1 there). In addition, data for two boundary layer measurements
are listed. Measurements were done by Knobloch and Fernholz at the Hermann-F\"ottinger Institute 
(HFI) and the German-Dutch Windtunnel (DNW) (for more details on the experiment, see Ref.~42). 
The current values listed here are calculated from a data analysis 
by Jachens.\cite{Jachens2004} The free-stream velocity 
for HFI is $U_{\infty}=10$ m/s and for DNW $U_{\infty}=80$ m/s. For the experiments, the shear
rate is given in 1/s and the kinematic viscosity of air is $\nu_{air}=1.5\times 10^{-5}$ m$^2$/s.}
\end{center}
\label{tab2}
\end{table}

%-----------------------------------------------------------------------

\section{Polymer stretching}

\subsection{Statistics of dumbbell extension}
Let us turn now to the stretching of the dumbbells in the described turbulent shear flow.
All parameters of the runs in Brownian dynamics are
summarized in Tab.~3. The corresponding turbulence parameters are as in Tab.~1.
According to Eq.~(\ref{dav1}), effects 
of mean shear will enter the Weissenberg number as well. 
A series of three simulations (2p, 3p, 4p) is conducted for which $\tau$ is
kept fixed and the Weissenberg number changes in correspondance with $\lambda_1(S)$ 
as shown in the upper panel of Fig.~3. In addition, runs 4p1 and  4p2 are at 
$Wi=0.26$ equal to that of 
run 2p, but at a larger shear rate of $S=5/\pi$. This series will demonstate in brief 
another possibility
of changing the physical parameters, namely keeping $S$ fixed and varying $Wi$ via the
relaxation time $\tau$. Furthermore, we will study the $R_0$ dependence of the 
extension statistics.  Our focus anyhow will be on the first series.
%-----------------------------------------------------------------------
\renewcommand{\arraystretch}{1.3}
\begin{table}
\begin{center}
\begin{tabular}{lcccccc}
\hline\hline
         & $S$     & $Wi$ & $N$ &  $N_d $          & $\eta/R_0$ & $R_0/\Delta$\\ \hline
run 2p   & $1/\pi$ & 0.26 & 512 &  $2\times 10^6$  & 5.5        & 2.0 \\
run 3p   & $3/\pi$ & 0.36 & 512 &  $2\times 10^6$  & 4.4        & 2.0 \\
run 4p   & $5/\pi$ & 0.59 & 128 &  $5\times 10^5$  & 1.7        & 1.0 \\ \hline 
run 4p1  & $5/\pi$ & 0.26 & 128 &  $5\times 10^5$  & 3.4        & 0.5 \\ 
run 4p2  & $5/\pi$ & 0.26 & 128 &  $5\times 10^5$  & 1.7        & 1.0 \\
\hline\hline
\end{tabular}
\caption{Parameters for the shear flow simulations with the dumbbell ensemble. Other parameters
of the simulations are as in Table~1. $N_d$ is the number of dumbbells and $N$ the
grid points in streamwise direction which corresponds to a resolution of 
$N_x\times N_y\times N_z=N\times(N/2+1)\times N$. We also list the ratio of the viscous Kolmogorov
scale to the equilibrium extension of the dumbbells and the ratio of the equilibrium extension to
the grid spacing. The grid spacing is $\Delta=2\pi/N$.}
\end{center}
\label{tab3}
\end{table}
%-----------------------------------------------------------------------  

\begin{figure}[htb]
\centerline{\includegraphics[angle=0,scale=0.2,draft=false]{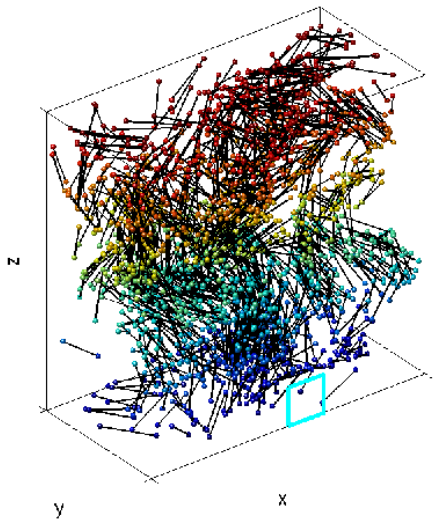}}
\caption{(color online) Instantaneous snapshot of the dumbbell distribution in the 
turbulent shear flow at $S=5/\pi$ and $Wi=0.26$ (run 4p2). Only the dumbbells with $R\ge 10\eta$
are plotted. The scale $10\eta$ is the sidelength of the square on the bottom
of the frontside of the box.}
\label{full}
\end{figure}
%---------------------------------------------------------------------------------------
  
In Fig.~5, we plot the instantaneous snapshot of 
all the dumbbells that are stretched beyond $10\eta$ for the shear rate $S=5/\pi$ (run 4p2). 
A first glance on the plot indicates no preferential orientation of the polymers and a 
rather 
complicated pattern in which they are arranged. This is exactly our motivation for a more 
detailed study of the angular statistics which will follow at the end of this section. 

First, we analyze the statistics of the end-to-end
norm $R$. Similar to isotropic flows we detect the occurance of the coil-stretch transition.
The numerical computation of the probability density function (PDF) $p(R,t)$
for $Wi<Wi_c=1/2$ and $Wi>Wi_c$ is shown in Fig.~6 and
Fig.~7 respectively. Note that the definition of the probability 
density function is chosen such that $\int_0^{\infty} p(R,t) dR =1$.
Runs 2p, 3p  and 4p2 correspond to the preferentially coiled state of the dumbbells, 
i.e. $Wi < 1/2$. The PDFs are stationary after a relaxation time that increases with 
$Wi$ and display fatter tails with increasing shear rate $S$. 
The dependence on the equilibrium extension $R_0$ is shown in the inset of Fig.~6. We find
that the PDF remains unchanged, except the very far tails. For both cases, $R_0$ was taken
well inside the regular and smooth viscous range of scales of the flow which makes our finding 
plausible.     

Run 4p corresponds with the preferentially stretched case for $Wi>Wi_c$. 
We plot a sequence of PDFs each taken at the times indicated in the legend of
Fig.~7. Although the small scale stretching is on average dominating the Hookean spring force
the PDF $p(R)$ reaches a stationary state.
One may loosely think of a mean field picture that the relative 
velocity between the beads scales as $\Delta{\bf u}(R)\simeq \lambda_1 R$.
Therefore under the condition that $\lambda_1 R \gtrsim R/(2 \tau)$, i.e. when the 
Weissenberg number is larger 1/2, the average inter-bead distance should grow with time.

As mentioned in the introduction, 
we determine the relative velocity between the beads by taking the {\em full}
increment between the individual bead velocities.
Consequently, the dumbbells get
stretched ultimately to scales $R \gtrsim \eta$ for $Wi>1/2$. At those scales the relative velocity between  
the beads scales as $\Delta{\bf u}(R)\sim R^{\alpha}$ with $0<\alpha<1$. 
Hence the growth of separation stops as the stretching term becomes
subdominant in comparison to linear relaxation by the spring force. 
We add two external scales of our system into the figure. The integral scale of turbulence 
$L_{int}$ (see Ref.~29 for definition) marks the average extension of the largest structures in the 
shear flow and
therefore the end of the inertial range. The box size $L_x$ is also shown. It can be observed that 
the sudden drop in the slope of $p(R)$ for
later times takes place at $R< L_x$, but at $R>L_{int}$. Note that the turbulent motion is 
completely decorrelated across scales $R>L_{int}$ and that large extensions (which go even beyond 
$R=L_x$) become more and more unprobable.     
  
Such a scenario is not new, it is a main ingredient of the cascade model of 
drag reduction in bulk turbulence suggested by Tabor and De Gennes.\cite{Tabor1986}
For a well established cascade with $\Delta{\bf u}(R)\simeq (\langle\epsilon\rangle R)^{1/3}$ 
a saturation scale of $R_*\simeq \,Wi^{3/2} \eta$ would follow. 
The mean extension $\langle R\rangle$ for the later plots of Fig.~7 is found 
to be at about 3$\eta$, i.e. the saturation takes place on average beyond the viscous scale of 
turbulence. In our case, the saturation is a result of
the nonlinear advection term in Eq.~(\ref{deq}) in combination with the Hookean spring
force.
%---------------------------------------------------------------------------------------
\begin{figure}[htb]
\centerline{\includegraphics[angle=0,scale=0.6,draft=false]{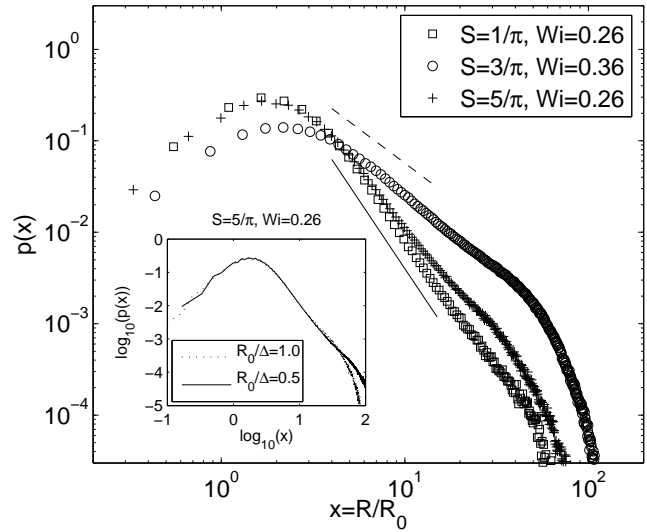}}
\caption{Probability density function (PDF) of the dumbbell extension $R$ as a 
function of the shear rate $S$. The length $R=|{\bf R}|$ is rescaled by the corresponding
equilibrium extension $R_0$ of the advecting shear flow. Symbols are indicated in the legend.
The slopes fitted to the tails are -3.0 for run 2p at $Wi=0.26$ (solid line) and -1.5 for 
run 3p at $Wi=0.36$ (dashed line). Run 4p2 at $S=5/\pi$ and $Wi=0.26$ is also 
shown in the main figure (see Tab.~3). Inset: $R_0$ dependence of the PDF of
dumbbell extension for two different equilibrium length values as indicated in
the legend (see also Tab.~3).}
\label{shearpdf}
\end{figure}
%---------------------------------------------------------------------------------------
\begin{figure}[htb]
\centerline{\includegraphics[angle=0,scale=0.6,draft=false]{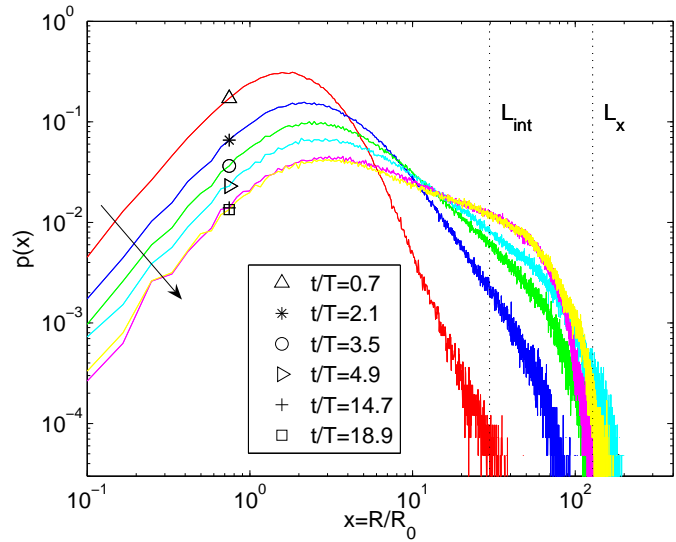}}
\caption{(color online) PDF of the dumbbell extension for the stretching 
at $S=5/\pi$ and $Wi=0.59$ (run 4p). The extension $R$ is rescaled again by the corresponding
equilibrium extension $R_0$. Curves at several times are plotted as indicated in the legend.
The inclined arrow to the left indicates growth in time.
$T=\langle v^2\rangle/(2\langle\epsilon\rangle)$ is the large scale eddy turnover time.
The vertical dotted lines are for the integral scale of turbulence $L_{int}$ and the box size 
$L_x=2\pi$.}
\label{tpdf}
\end{figure}
%-----------------------------------------------------------------------

\subsection{Anisotropy of stretching}
It is important to observe that the effect of shear cannot be merely 
translated to an effective scaling of the Weissenberg number. 
The shear flow is not isotropic and hence the stretching shall not be so.
To track the impact of shear we have plotted in Fig.~8 the PDFs 
of the individual components $R_i$ for two cases below $Wi=1/2$. 
As expected, the stretching in the streamwise direction becomes dominant when
the Weissenberg number increases.   The extension in $y$ direction is always bounded to $L_y$ 
due to the free slip boundary conditions. This effect is also existing for very 
low shear rates. 

Figure~9 shows the second order moments 
$\langle R_i R_j \rangle$ for run 3p at $Wi=0.36$. Effects of anisotropy are well 
detectable in the figure. The moment $\langle R_x R_y \rangle$ has a finite positive 
value while the other two mixed moments were found to vary around zero.
%-----------------------------------------------------------------------
\begin{figure}[htb]
\centerline{\includegraphics[angle=0,scale=0.6,draft=false]{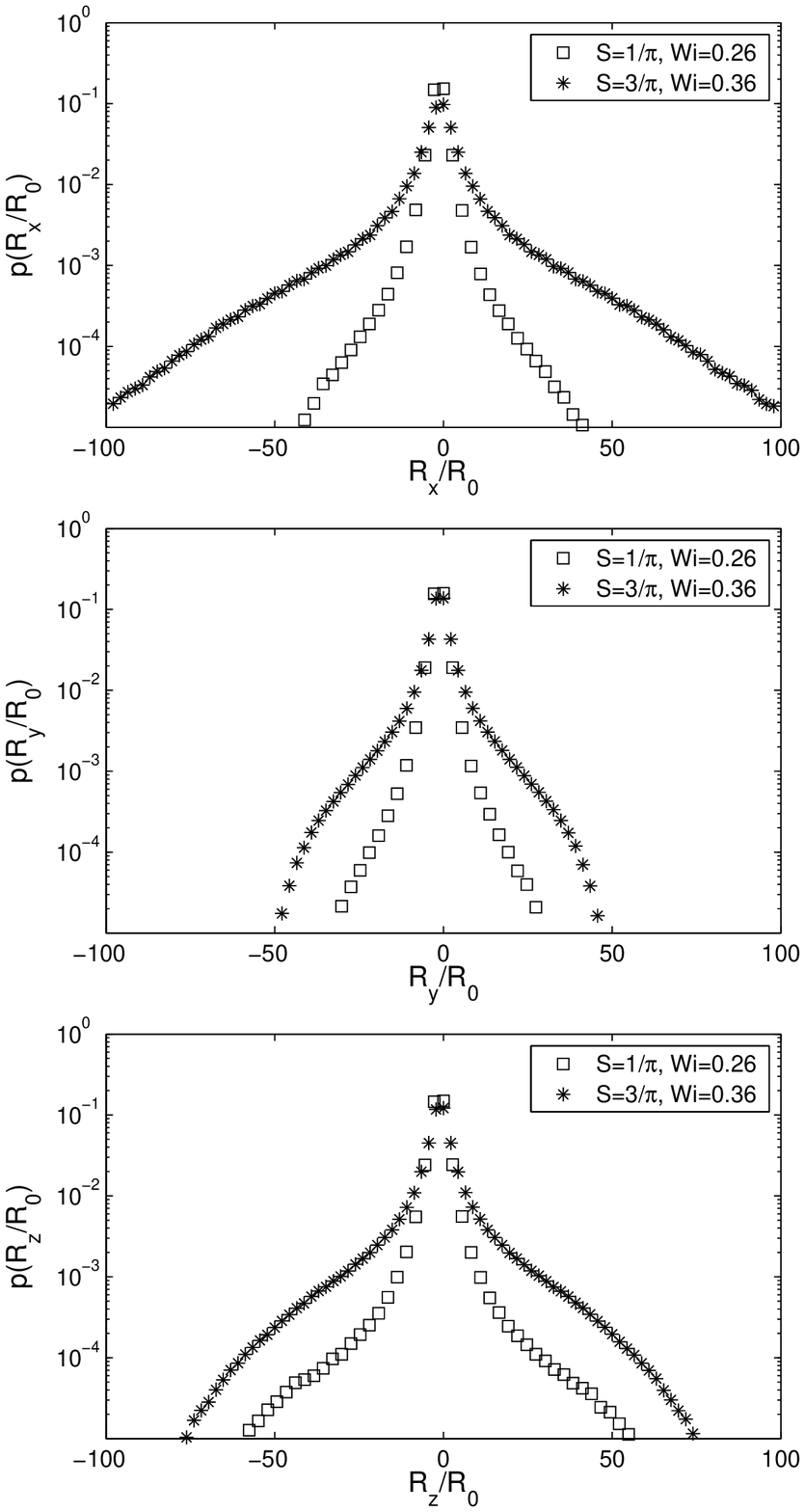}}
\caption{Anisotropy of the stretching as quantified by the PDFs $p(R_i/R_0)$
for the streamwise component $i=x$ (upper panel), the shear component $i=y$ (mid panel), and
the spanwise component $i=z$ (lower panel).}
\label{shearpdf_c}
\end{figure}
%-----------------------------------------------------------------------
Moreover, the moments $\langle R_x^2 \rangle$ , $\langle R_y^2 \rangle$ and 
$\langle R_z^2 \rangle$ are no longer collapsing as for a lower shear rate. The stretching 
in the streamwise direction gives not only the largest extensions, but also the largest 
fluctuations. 
When comparing it with a graph of the turbulent kinetic energy of the corresponding run 3p 
(see inset of Fig.~9), we recognize similar patterns of time variation. They 
indicate that the large scale variations of the turbulence fluctuations have an effect on the 
stretching history, most dominantly in streamwise $x$-direction. For that reason we
have to take statistical averages over longer time intervals with increasing $S$. 

Such fluctuations are known for homogeneous and nearly homogeneous shear flows.
We recall that the scale of variation of the mean velocity gradient is outside 
the considered simulation domain. Simulations which start with isotropic initial
conditions will show a self-similar growth of integral scale and  turbulent kinetic 
energy until length scales of the finite simulation domain are reached. A statistically 
stationary regime is then accompanied by fluctuations of the turbulent kinetic energy. 
\cite{Gualtieri2002,Yakhot2003,Schumacher2004}  Such velocity (and kinetic energy) 
fluctuations show up
as streamwise streaks in the shear flow. Our studies are also in agreement
with recent works by Stone and Graham\cite{Graham2003} and Terrapon {\it et al.}\cite{Terrapon2004} 
who found preferential stretching near streamwise streaks. Both cases considered a so-called 
minimal 
flow unit that captures important features of a turbulent buffer layer. In the latter work such 
preferential stretching study was detected by an eigenvalue analysis of the local velocity gradient
along the dumbbell tracks.
The streaks decay due to linear instabilities
and become less coherent when the Reynolds number increases. \cite{Schumacher2001}
We expect that such pronounced variations in the streamwise extensions will decrease for
larger Reynolds numbers.

%-----------------------------------------------------------------------
\begin{figure}[htb]
\centerline{\includegraphics[angle=0,scale=0.6,draft=false]{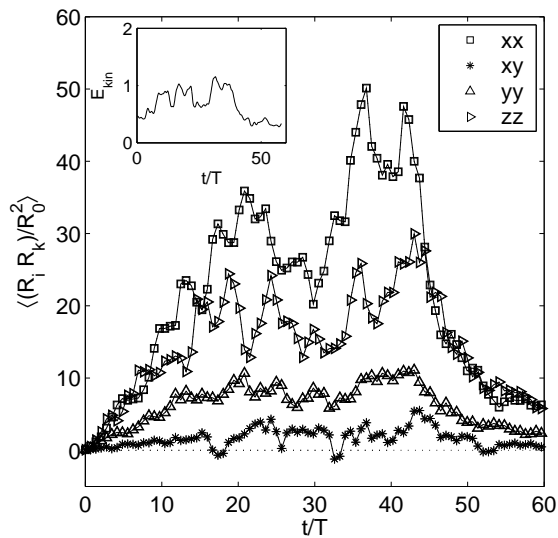}}
\caption{Time evolution of the polymer conformation tensor components 
as quantified by $\langle R_i R_k\rangle/R_0^2$. Data are for $S=3/\pi$ and $Wi=0.36$. 
The symbols of the four components 
are given in the legend. Times are rescaled by the large scale eddy turnover
time $T=\langle v^2\rangle/(2\langle\epsilon\rangle)$. 
Inset: Turbulent kinetic energy as a function of dimensionless time $t/T$. 
The long time analysis was conducted at $N=128$.}
\label{shear2}
\end{figure}
%----------------------------------------------------------------------

\subsection{Angular statistics}
The results of the last chapter suggest an investigation of the statistics of 
the azimuthal angle $\phi$ which measures the orientation of the dumbbells with respect 
to the mean flow component. The azimuthal angle is calculated as (see also Fig.~1)
%-----------------------------------------------------------------------------
\begin{equation}
\phi=\mbox{arctan}\left(\frac{R_y}{R_x}\right)\;, 
\end{equation}
%-----------------------------------------------------------------------------
with values between $-\pi/2$ and $\pi/2$. Figures~10 and 11 show our 
analysis for the runs 2p, 3p and 4p (see also Tab.~3). 
To resolve the anisotropy of stretching,
the contour plot of the joint PDFs of extension and 
azimuthal angle, $p(R,\phi)$, is numerically computed and plotted in Fig.~10.
They are normalized such that $\int_0^{\infty} dR \int_{-\pi/2}^{+\pi/2} d\phi~p(R,\phi)=1$.
The lowest shear rate at $S=1/\pi$ shows practically no variation
with respect to $\phi$ (see upper panel of Fig.~10). 
For highly stretched dumbbells the $\phi$ probability distribution 
is essentially flat for almost all $\phi$ values exept a narrow region around 
$\phi\simeq 0$. However this tendency of alignment with the mean flow is very weak and only 
visible for the most extended polymers. The dumbbell dynamics is very similar to that
of an isotropic flow.

The picture changes in case of higher shear rates as can be seen in the lower
panel of Fig.~10. We observe a larger number of dumbbells aligned 
around the mean flow axis at $\phi=0$ even when they are not very extended. Furthermore, in 
comparison with lower shear, it is less probable to observe 
dumbbells perpendicular to $x$-axis, i.e. at $\phi=\pm\pi/2$. A closer inspection of 
both plots unravels an asymmetry in $\phi$ probability distribution.
Dumbbells with
$\phi\in [-\pi/2,0]$ are unstable with respect to the mean flow in contrast to those with 
$\phi\in [0,\pi/2]$.  
%----------------------------------------------------------------------
\begin{figure}[htb]
\centerline{\includegraphics[angle=0,scale=0.6,draft=false]{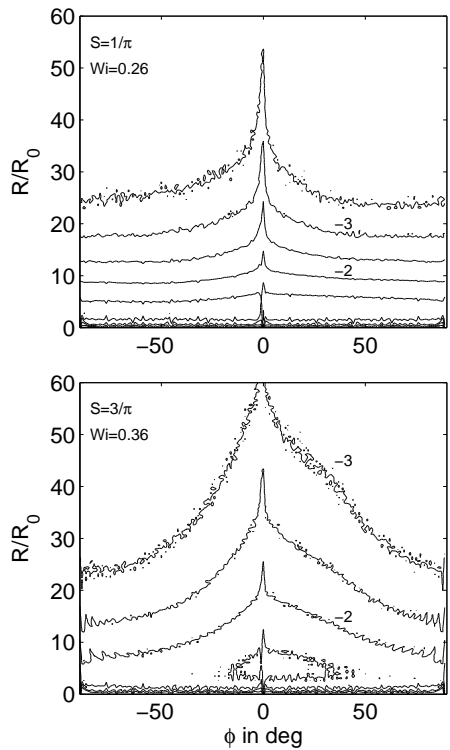}}
\caption{Joint probability density function $p(R/R_0,\phi)$ for two
different simulations. The contour levels of $p(R/R_0,\phi)$ are given in units of
the decadic logarithm and decrease in steps of 0.5. 
Upper panel: data are for run 2p with a shear rate $S=1/\pi$ and $Wi=0.26$.
Lower panel: data are for run 3p with a shear rate $S=3/\pi$ and $Wi=0.36$. 
The contour lines for $10^{-3}$ and $10^{-2}$ are indicated by exponents -3 and -2, respectively.}  
\label{angle}
\end{figure}
%-----------------------------------------------------------------------
\begin{figure}[htb]
\centerline{\includegraphics[angle=0,scale=0.6,draft=false]{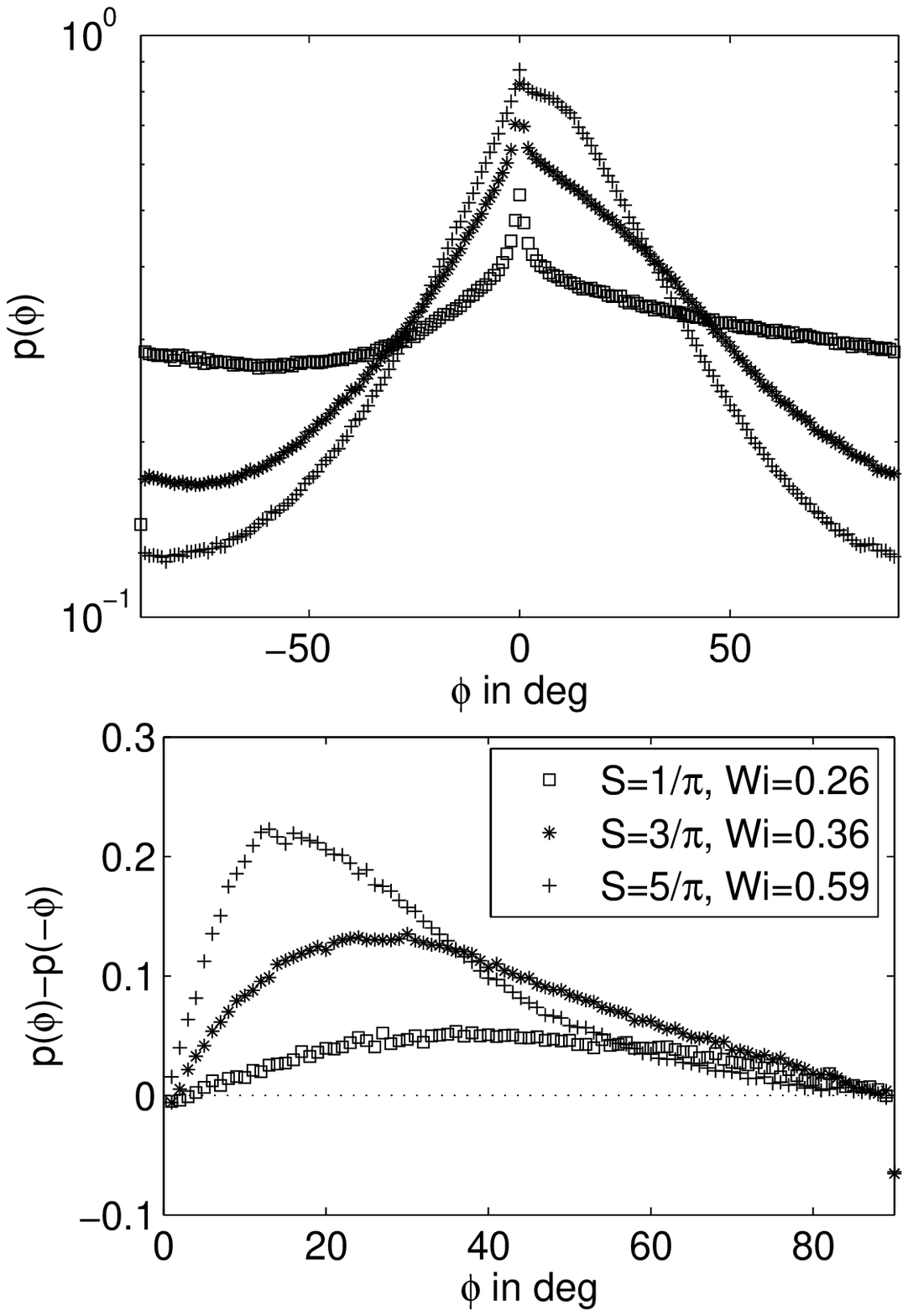}}
\caption{Upper panel: probability density function $p(\phi)$ for 
the three runs 2p, 3p, 4p as given in Tab.~3. Lower panel: Corresponding asymmetry 
of $p(\phi)$ as given by relation (\ref{asymmetry}).} 
\label{angle1}
\end{figure}
%----------------------------------------------------------------------
 
The PDF of the azimuthal angle is calculated by integration of the joint PDF
%--------------------------------------------------------------
\begin{equation}
p(\phi)=\int_0^{\infty}\mbox{d}R\,p(R,\phi)\,.
\end{equation}
%--------------------------------------------------------------
The angular distribution is shown in the upper panel of Fig.~11. The asymmetry between both
quadrants is quantified by the following measure
%--------------------------------------------------------------
\begin{equation}
A(\phi)=p(\phi)-p(-\phi)\,,
\label{asymmetry}
\end{equation}
%--------------------------------------------------------------
with $\phi\in[0,\pi/2]$. The measure $A(\phi)$ is plotted in the lower panel of Fig.~11.
Note that the integral of $A(\phi)$ determines the {\it total} asymmetry of the PDFs.
We find that with increasing shear rate the asymmetry of the angular
distribution grows. The plot of $p(\phi)$ shows a growing peak around $\phi=0$ with increasing 
$Wi$. Fluctuations in the vicinity of $\phi=0$ are enhanced while the tails for 
very large $\phi$ are depleted. 

%----------------------------------------------------------------------
\begin{figure}[htb]
\centerline{\includegraphics[angle=0,scale=0.6,draft=false]{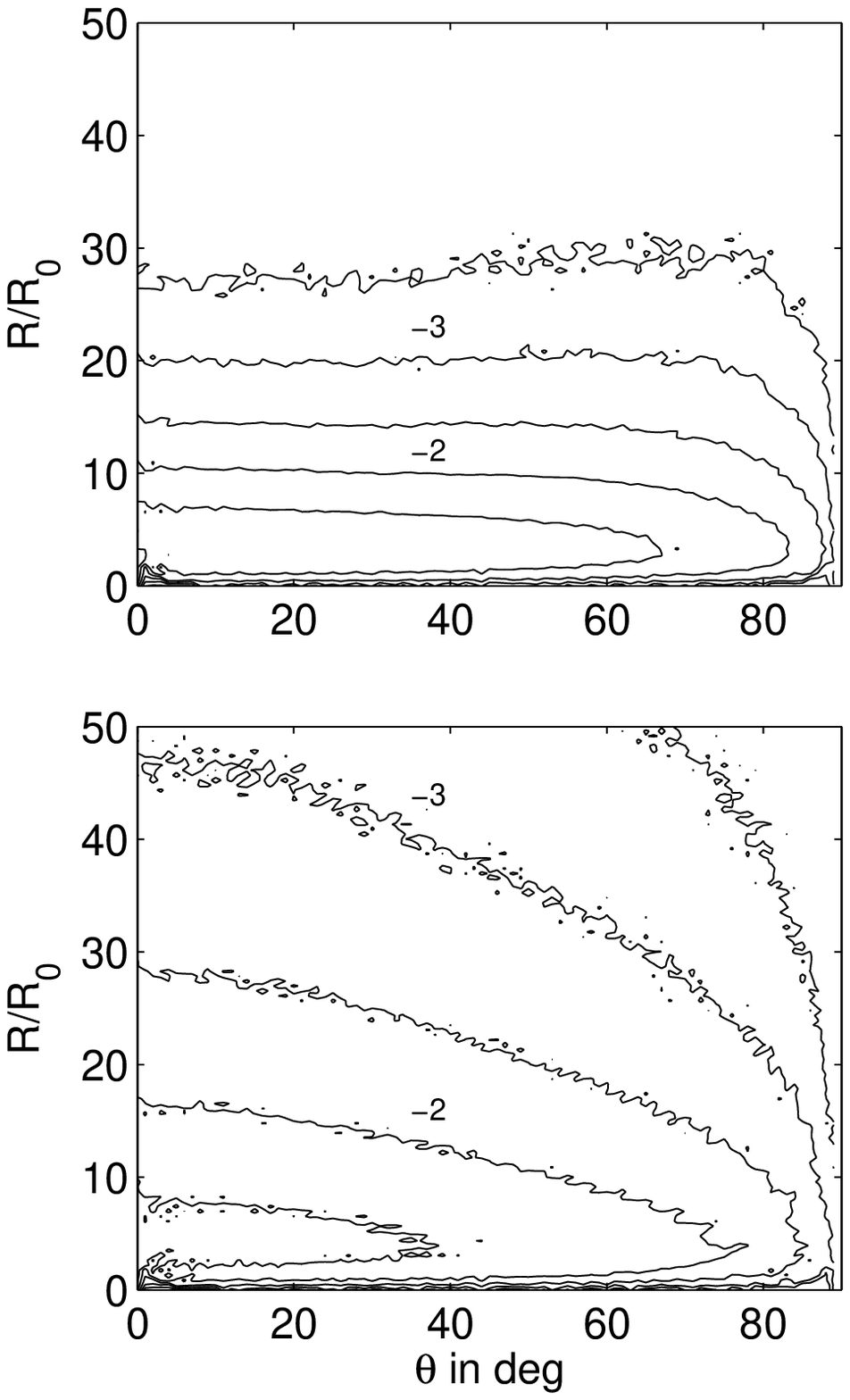}}
\caption{Joint probability density function $p(R/R_0,\theta)$ 
for two different simulations. Data and line spacing are the same as for Fig.~10.
Exponents for levels $10^{-3}$ and $10^{-2}$ are indicated again.}
\label{polar}
\end{figure}
%-----------------------------------------------------------------------
The second angular degree of freedom is the polar angle defined as,
\begin{equation} 
\theta=\mbox{arctan}\left(\frac{R_z}{\sqrt{R_x^2+R_y^2}}\right),
\end{equation}
where $\theta \in [-\pi/2,\pi/2]$.
The statistics of $\theta$ is therefore reconstructed by having the full information
of the three components $R_i$.
The simultaneous statistical information of end-to-end norm $R$ and the polar angle $\theta$
is expressed by the joint probability distribution $p(R,\theta)$ which is normalized as
$\int_0^{\infty} dR \int_{-\pi/2}^{+\pi/2}
d\theta~p(R,\theta)=1$. 

Figure~12 shows the contour plots of the joint distribution $p(R,\theta)$ for runs 2p and 3p. 
Now, $\theta \in [0,\pi/2]$ because the dynamics is invariant under the transformation
$\theta \to -\theta$ (which we verified in our simulations).
The probability distribution of $R$ at a fixed polar angle is given by vertical cuts
in the two-parameter plane.
For $S=1/\pi$ (upper panel of Fig.~12), the contour levels in the 
vertical direction are very close and insensitive
to the polar angle. A low Weissenberg number at this shear rate is consistent with close
spacing of the contours in $R$ direction and indicates a rapid decrease of the extension 
probability. Moreover, the decay of $p(R,\theta)$ with $R$ at fixed $\theta$ is more or less similar
for the majority of polar directions.
Recall that a larger polar angle $\theta$ means that dumbbells are less aligned with the $x$-$y$ 
shear plane. Consequently, orientations aligned with the shear plane and off that plane are equally
probable.
For $S=3/\pi$ (lower panel of Fig.~12), the contour level spacing in the $R$ direction 
is larger, obviously a result of the increase of Weissenberg number.
The other feature is that the projected distributions along the $R$ direction decrease more rapidly 
with increasing the polar angle $\theta$.
This manifests the increase of the anisotopy in $\theta$ direction. Therefore the 
dumbbells aligned with $x$-$y$ shear plane are highly stretched only.

To demonstrate this effect, an instantaneous snapshot of the most extended 
dumbbells for run 4p2 is shown in Fig.~13. Again, dumbbells with $R\ge 10\eta$ are 
plotted only, but now in two different sectors of the polar angle. 
A preferential alignment of the dumbbells with smaller polar 
angles - sector $0\le \theta\le 3\pi/20$ is compared with sector $7\pi/20\le \theta\le \pi/2$ -
can be clearly detected from both panels (see also Fig.~5).

Figure~14 shows the probability distribution of polar angle,
%--------------------------------------------------------------
\begin{equation}
p(\theta)=\int_0^{\infty}\mbox{d}R\,p(R,\theta)\,.
\end{equation}
%--------------------------------------------------------------
The comparison of run 4p with runs 2p and 3p shows that the number of dumbbells  
aligned with the $x$-$y$ plane increases with shear rate. This is reflected in
the increasing maximum of the probability distribution around $\theta 
\approx 0$.   
%-----------------------------------------------------------------------
\begin{figure}[htb]
\centerline{\includegraphics[angle=0,scale=0.1,draft=false]{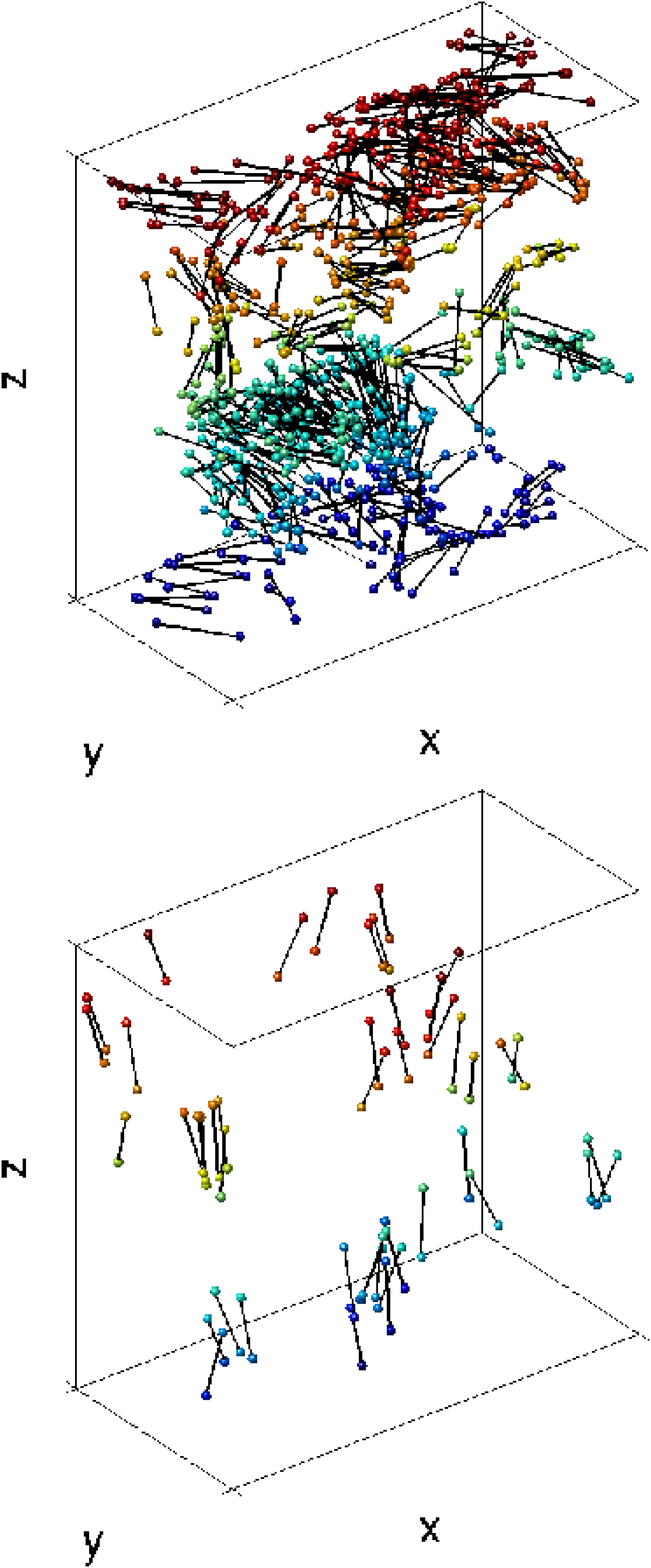}}
\caption{(color online) Instantaneous snapshot of the dumbbell distribution in the 
turbulent shear flow (run 4p2). Same data as in Fig.~5 are used. Upper panel: dumbbells
for $R\ge 10\eta$ and $|\theta|<27$ degrees. Lower panel:  dumbbells
for $R\ge 10\eta$ and $|\theta|>63$ degrees.}
\label{full1}
\end{figure}
%-----------------------------------------------------------------------
\begin{figure}[htb]
\centerline{\includegraphics[angle=0,scale=0.6,draft=false]{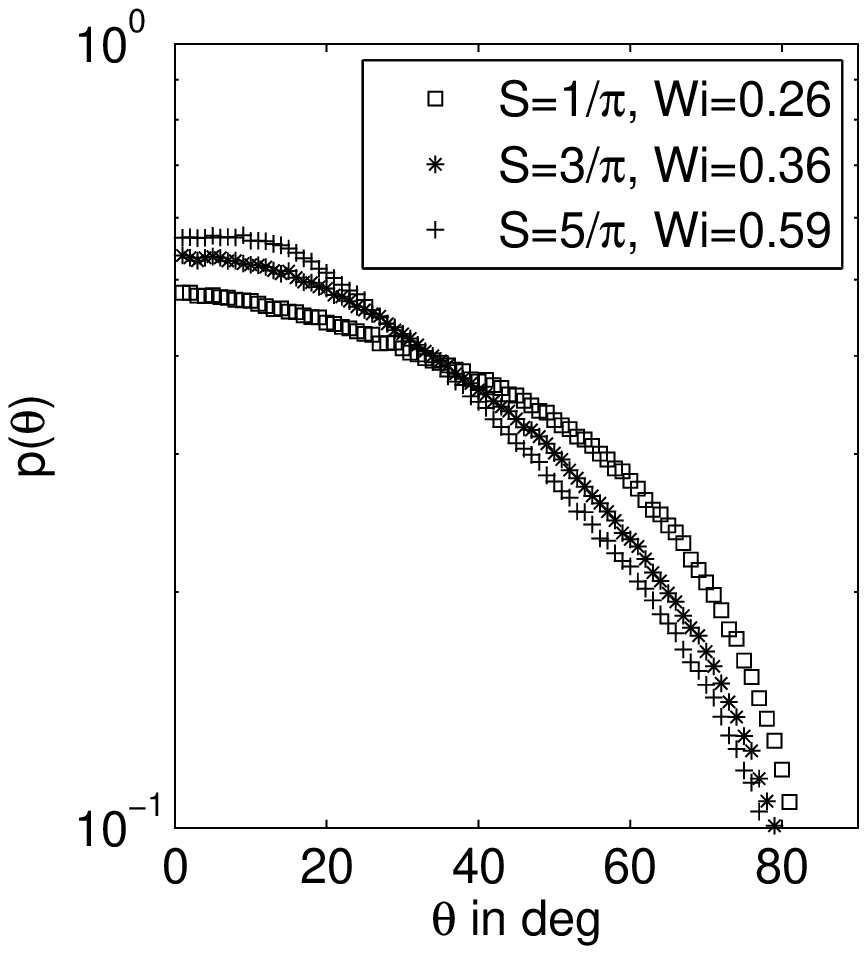}}
\caption{Probability density function $p(\theta)$ for the three  
runs 2p, 3p, 4p as given in Tab.~3.}
\label{angle2}
\end{figure}
%-------------------------------------------------------------------------------

\section{Summary and discussion}
We present numerical studies of the stretching of Hookean dumbbells in turbulent 
flows with different mean shear rates which are described by the Navier-Stokes dynamics. 
The flow at hand is the simplest statistically stationary turbulent shear flow, a nearly
homogeneous shear flow with a linear mean profile. It can be thought as the flow in the
bulk of a turbulent shear flow with no-slip boundaries, e.g. in a turbulent plane Couette flow.  
Beside the turbulence structures above the Kolmogorov scale $\eta$,
the dumbbell dynamics around their equilibrium distance $R_0<\eta$ is well resolved. The latter 
is important for the lower Weissenberg numbers where the majority of dumbbells remain below
the viscous scale. This constraint clearly limits the accessible range of Reynolds numbers. 

In the first part, we 
studied the stretching properties of the flow that are quantified by the largest
finite-time Lyapunov exponent. For larger values of $S$ our results infer that
$\lambda_1\sim S^{3/2}$ which is consistent with a simple 
dimensional analysis. Our studies and further supplementary
data from other shear flows show that the Corrsin parameter does not get
much larger than unity (in fact our data are all below 1). This point is important 
since it determines which flow time scale has to be compared with the
relaxation time of the polymer chains. In our study the relevant time scale is 
$\sim\tau_{\eta}$, in contrast to
$S^{-1}$ as in a recent analytical model.\cite{Chertkov2005}

In the second part of the work, we studied the shear rate dependence of the stretching
properties of Hookean dumbbells. The growth of anisotropy with shear rate is 
quantified by measuring the statistics of $R_i$, the azimuthal angle
$\phi$, and the polar angle $\theta$. When $S$ grows
a preferential orientation at azimuthal angles that are slightly larger than $\phi=0$ 
is observed. Moreover 
the probability distribution of the polar angle shows a preference for smaller values of 
$\theta$ with increasing the shear rate. 
Therefore more polymers align with the shear plane as the shear rate gets higher.
When keeping the shear rate fixed, the alignment of 
the dumbbells with the shear direction depends on the magnitude of the relaxation time $\tau$. 
For higher Weissenberg numbers and therefore larger relaxation times a more pronounced anisotropic 
stretching is then observed. Furthermore, the $p(R)$ is found to be insensitive with respect to a 
variation of the equilibrium length $R_0$ for most extension scales.

Streamwise streaks which are a characteristic feature of shear flows are found
to cause variations in the polymer extension fluctuations. In agreement with previous
studies for the case of a minimal flow unit of a turbulent boundary 
layer\cite{Graham2003,Terrapon2004}, these structures are found to enhance the stretching.    
For $Wi >1/2$ the Hookean potential the inter-bead separations extend to
the spatial scales where they experience a rough relative velocity difference. 
Therefore we detect a saturation of the stretching, 
even for Weissenberg numbers beyond 1/2 and in agreement with cascade models 
for drag reduction in the bulk of turbulence.\cite{Tabor1986}

It will be interesting to investigate the behavior of finitely extensible nonlinearly elastic 
(FENE) dumbbells in our flow configuration. A further parameter, the maximum length 
$L_0$ of the dumbbells, will have then an impact on the results. 
In that case well resolved velocity data below the Kolmogorov 
scale become very important. This is due to the fact that even for the largest polymer chains in 
turbulent flows $L_0\lesssim\eta$. These studies are part of our future work.

\acknowledgements
The work is supported by the Deutsche Forschungsgemeinschaft (DFG) within the Interdisciplinary
Turbulence Initiative and by the German Academic Exchange Service (DAAD) within the PROCOPE
program. J.S.
wishes to acknowledge partial support by the NSF and thanks E. Bodenschatz for his
hospitality at Cornell University where parts of this work were done. Discussions
with J. Bec, E. Bodenschatz, A. Celani, L. R. Collins, C. R. Doering, B. D\"unweg, B. Eckhardt, X. Hu, 
N. Ouellette, A. Puliafito, K. R. Sreenivasan and 
D. Vincenzi are acknowledged. The computations were carried out on the IBM JUMP
Cluster at the John von Neumann-Institute for Computing of the Research Center J\"ulich (Germany).
We want to thank for their steady support with computing ressources.

\end{document}